\documentclass[floatfix, aps,a4,twocolumn,prd,preprint,10pt,superscriptaddress]{revtex4}

\usepackage{graphicx}

\begin{document}

\title{Ground states and dynamics of a trapped charged particle in the magnetic field}

\date{\today}

\author{Maciej Janowicz and Jan Mostowski}
\affiliation{Institute of Physics of the Polish Academy of Sciences,
Aleja Lotnik\'ow 32/46, 02-668 Warsaw, Poland}

\begin{abstract}
A system of two charged particles in a harmonic trap with additional
magnetic field is considered. The problem is reduced to a single-particle
one in relative coordinates. The ground- and lowest excited-state energies
and wave functions are found. The ground state exhibits non-zero expectation
value of the velocity (kinetic momentum) and the probability current
density does not vanish as well. When the ground state becomes degenerate the expectation
value of velocity becomes discontinuous. The effects associated with turning on
of the magnetic field are studied by solving the appropriate
time-dependent Schr\"odinger equation. No substantial differences between
abrupt (discontinuous in time) and continuous switching on have
been observed. Evolution of a wave packet which is initially Gaussian
is also investigated. The wave packet loses its Gaussian nature
and, after sufficiently large time, a system of diffractive maxima
and minima is built.
\end{abstract}

\keywords{trapped particles in magnetic field; kinetic momentum; wave packet evolution; degenerate ground state}

\pacs{03.65.Ge; 31.15.-p}

\maketitle


\section{Introduction}
\label{Intro}

Time crystals and their dynamics became an interesting problem after recent publications of \cite{Wilczek},
etc. In short time crystals are systems where natural
symmetry is not compatible with gauge field. A good example is a system of two or more ions in 
a cylindrically symmetric harmonic trap and 
in an additional constant homogeneous magnetic field directed along the symmetry axis. Because of cylindrical 
symmetry in the ground state the system wave function should be an eigenstate of the angular momentum component
along the symmetry axis. On the other hand in is a property of the magnetic field that the total flux 
of the field should be a multiple of the flux quantum. These two conditions are not compatible with 
each other. It has been argued \cite{Wilczek}that the ground state as well as excited state 
of the system show a nontrivial dynamics. In particular the ground state of a time crystal is not stationary, 
but instead exhibits time dependence characteristic for motion. We will explore this idea in detail in this paper.

A system of ions on a trap with additional magnetic field was recently 
investigated by \cite{LGYQYZDZ} from the point of view of time crystal dynamics. 
It was shown there that two o more ions indeed provide a good example o a time crystal and the dynamics 
was discussed.

In this paper we will further investigate similar systems. In particular we will discuss the 
dynamics, i.e. time dependence of the wave function. We will restrict ourselves to the simplest 
case of two ions, this simple system exhibits all the characteristic features of a time crystal. More ions 
add complications not giving any new aspects to the problem.
there is no need to discuss more ions.

The problem of interaction of two charged non-relativistic particles in a trap
with added homogeneous magnetic field is interesting also from another point of view. This is one of
the simplest systems in which non-trivial effects associated with the difference
between the canonical and kinetic momentum can be studied.
The particular form of the coupling of the electromagnetic field to the matter
fields appears as a consequence of the gauge invariance of quantum
electrodynamics. Even in the realm of non-relativistic quantum mechanics
that coupling leads quite profound effects, in particular, to the necessity
to distinguish between the canonical and kinetic momentum. In the presence of
(non-static) electromagnetic field the canonical momentum while retaining its role
in the Hamiltonian formalism loses its status of a physical observable and
is to be replaced with the velocity. This happens even in the simplest
situation when the electromagnetic field reduces itself to the constant homogeneous
magnetic field when the corresponding vector potential is linear in the coordinate.

One may suspect that a system with non-zero expectation value of the radial
coordinate (hence non-zero expectation value of the vector potential) can
be of particular interest. Indeed, in such a system the expectation value
of velocity does not vanish in the ground state because the minimum
of the effective potential is displaced from zero. One way to achieve that
is to introduce a repulsive Coulomb potential with the center at the same
point as the center of the trap potential. This can be realized by placing
to ions with identical charges in a trap. Then the problem reduces to a one-body
problem in the relative coordinates.
Similar ring-shaped traps have been considered, e.g., in
\cite{LGYQYZDZ,SSH,MG,OYTWSO,CMPHKK,Clark}. 
A remarkable comparison of the energy spectra as produced by the propagation
of a semiclassical wave packet and those computed from the WKB approximation
has been performed in \cite{GK}.
A real and very interesting related
systems are composed of aromatic molecules in the magnetic field \cite{GM,MHS}

The main part of this work is organized as follows. In Section 2, the construct
the mathematical model and describe the basis states. Section 3 contains
the results of the calculation of lowest energy levels as well as teh
ground-state expectation value of the velocity. In Section 4 we investigate
the effects associated with abrupt turning on of the magnetic field.
Section 5 is devoted to the evolution of the wave packet, and
Section 6 contains some concluding remarks.

\section{The model}

Let us consider a system of two identically charged particles which are trapped
in a harmonic external potential with additional homogeneous magnetic field.
The Hamiltonian of one of the particles in the relative coordinates
(in two spatial dimensions) can be written as:

\begin{equation}
H = \frac{1}{2 \mu} \left( {\bf p} - e {\bf A} \right)^{2} + \frac{1}{2}
\mu \omega_{t}^{2} {\bf r}^{2} + \frac{k e^{2}}{|{\bf r}|},
\end{equation}

where $\mu$ is the reduced mass, $e$ is the charge of each particle,
${\bf p}$ is the relative canonical momentum,
${\bf A}$ is the vector potential, $\omega_{t}$ is the frequency of the trap,
$k = 1/(4 \pi \epsilon_{0})$, and ${\bf r} = (x, y)$ are the relative coordinates.
We choose the so-called symmetric gauge for the vector potential to represent
the homogeneous magnetic field (which sets the $z$-axis of the system):

$$
{\bf A} = \frac{1}{2} {\bf B} \times {\bf r},
$$

where ${\bf B} = (0, 0, B_{z})$ is the magnetic induction.

Before we proceed let us briefly discuss the motion of particles within the framework 
of classical mechanics. It is clear that the lowest energy solution of the classical
equations of motion are obtained when the velocities  
$\frac{1}{\mu} \left( {\bf p} - e {\bf A} \right)$ equal to zero. Thus there is no 
motion if the system is in the lowest energy configuration. The canonical momentum
${\bf p}$ is different from zero, but this does not relate to movement.

We will now proceed with description of the system within the framework of quantum mechanics. 
Let us introduce the dimensionless coordinates $\xi$, $\eta$ such that

$$
x = \frac{\sqrt{\hbar}}{\sqrt{\mu \omega_{t}}} \xi, \;\;
y = \frac{\sqrt{\hbar}}{\sqrt{\mu \omega_{t}}} \eta.
$$

Then the Hamiltonian takes the form:

\begin{eqnarray}
H &=& {\hbar} \omega_{t} \left( -\frac{1}{2} \frac{\nabla_{\rho}^{2}}{2}
+ (\frac{1}{2} + \frac{1}{8} \nu^{2}) \rho^{2} + \right. \nonumber \\
&+& \left. \frac{i}{2}(\nu (\xi \partial_{\eta} - \eta \partial_{\xi}) + b/\rho
\right),
\end{eqnarray}

where $\rho = \sqrt{\xi^{2} + \eta^{2}}$,
$\partial_{\xi} = \partial/\partial_{\xi}$,
$\partial_{\eta} = \partial/\partial_{\eta}$, $\nabla_{\rho}^{2} =
\partial_{\xi}^{2} + \partial_{\eta}^{2}$, $\nu = \omega_{c}/\omega_{t}$,
$b = (k e^{2}/{\hbar}) \sqrt{(m/({\hbar} \omega_{t}))}$,
and $\omega_{c} = e B_{z}/m$ is the cyclotron frequency.
Thus, the parameter $\nu$ is a measure of strength of the magnetic field.
The dynamics of the system are determined by the two parameters $\nu$ and $b$.

In polar coordinates $\rho, \phi$ such that $\xi = \rho \cos{\phi}$,
$\eta = \rho \sin{\phi}$ the Hamiltonian takes the form:

\begin{eqnarray}
H &=& \hbar \omega_{t} \left[-\frac{1}{2} \left( \frac{\partial^{2}}{\partial \rho}^{2}
+ \frac{1}{\rho} \frac{\partial}{\partial \rho} \right) + \right. \nonumber \\
&+& \left. \frac{i}{2} \nu \frac{\partial}{\partial \phi}
+ \frac{1}{2} (1 + \frac{1}{4} \nu^{2}) \rho^{2} - \frac{1}{2} \frac{1}{\rho^{2}}
\frac{\partial^{2}}{\partial \phi^{2}}
+ \frac{b}{\rho} \right].
\end{eqnarray}

Clearly, the $z$-component of the angular momentum commutes
with the Hamiltonian, and the latter admits separation of variables
in the radial coordinates $\rho$ and $\phi$.

Let us consider the time-independent Schr\"odinger equation

\begin{equation}
H \psi = E \psi,
\end{equation}

and let us write the wave function $\psi = \psi(\rho, \phi)$
in terms of separated variables ($\rho$, $\phi$) as:

\begin{equation}
\psi(\rho, \phi) = \frac{1}{\sqrt{\rho}} \chi(\rho) e^{i m \phi}.
\end{equation}

Then the time-independent Schr\"odinger equation takes the form:

\begin{equation}
\label{TIprime}
H^{\prime} \chi(\rho) = E \chi(\rho),
\end{equation}

where

\begin{equation}
H^{\prime} = \frac{1}{2} {\hbar} \omega_{t} \left[
-\frac{\partial^{2}}{\partial \rho^{2}} - m \nu + (1 + \frac{\nu^{2}}{4}) \rho^{2}
+(m^{2} - \frac{1}{4}) \rho^{-2} + \frac{b^{\prime}}{\rho} \right],
\end{equation}

where $b^{\prime} = 2 b$. $H^{\prime}$ can also be written as a sum of the kinetic
term and dimensionless effective potential term:

$$
H^{\prime} = \frac{1}{2} {\hbar} \omega_{t}
\left(-\frac{\partial^{2}}{\partial \rho^{2}} + V(\rho) \right).
$$

With the above forms of the Hamiltonian we are well-prepared to study the spectral
and dynamical features of the system.

In Fig. 1 we have displayed the effective potential $V(\rho)$ as a function
of $\rho$.

\begin{figure}
\begin{center}
\includegraphics[width=7cm, height=6cm]{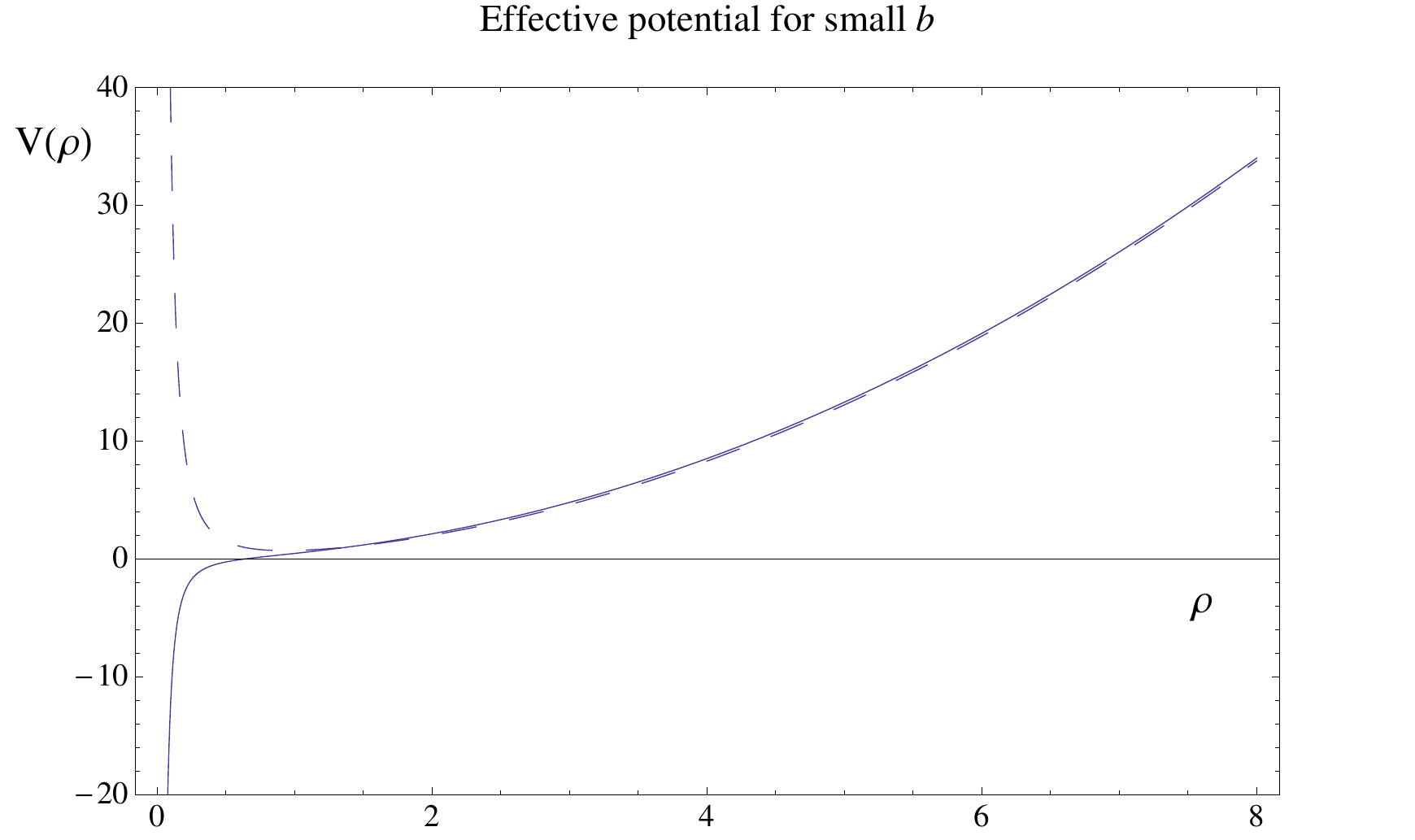}
\includegraphics[width=7cm, height=6cm]{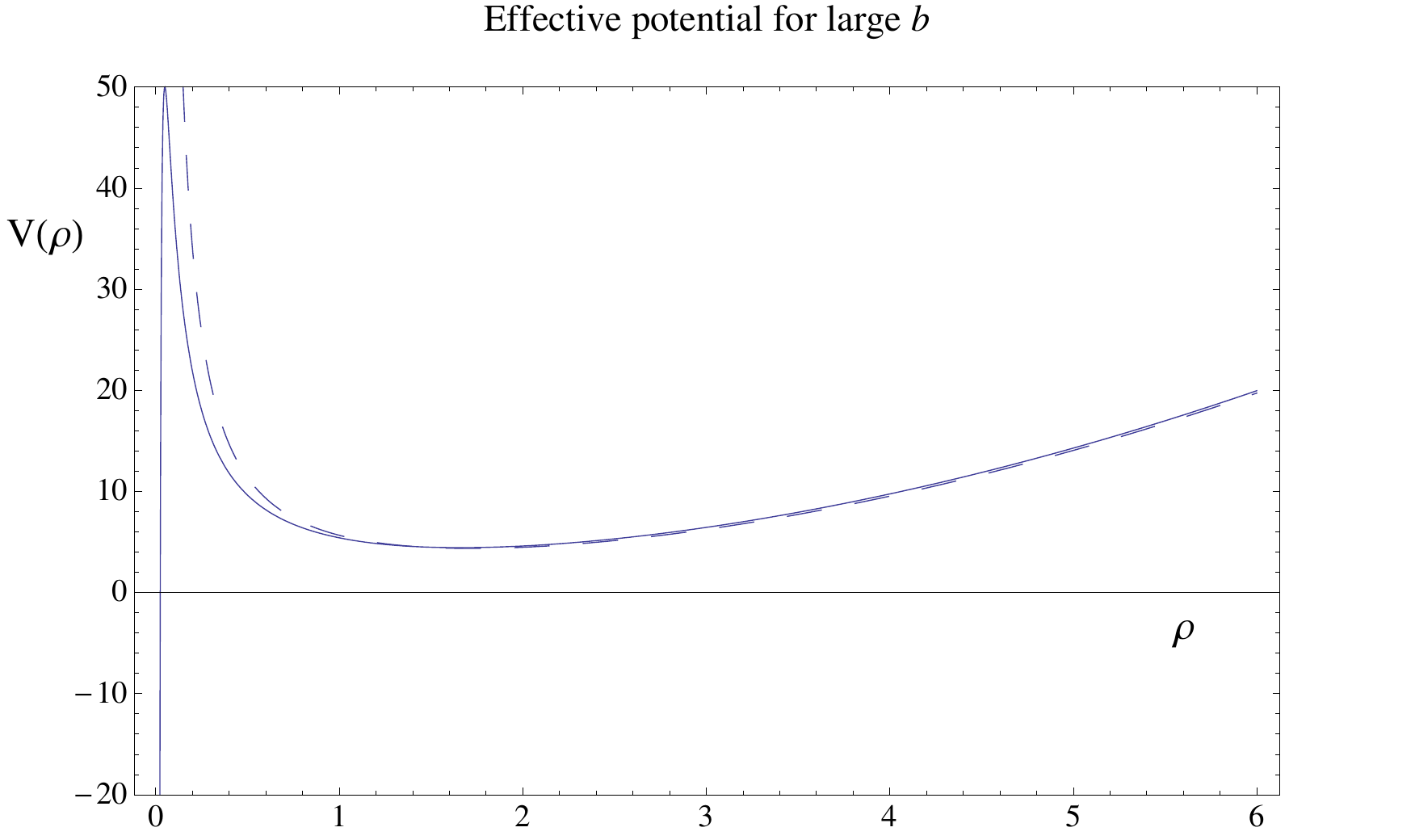}
\includegraphics[width=7cm, height=6cm]{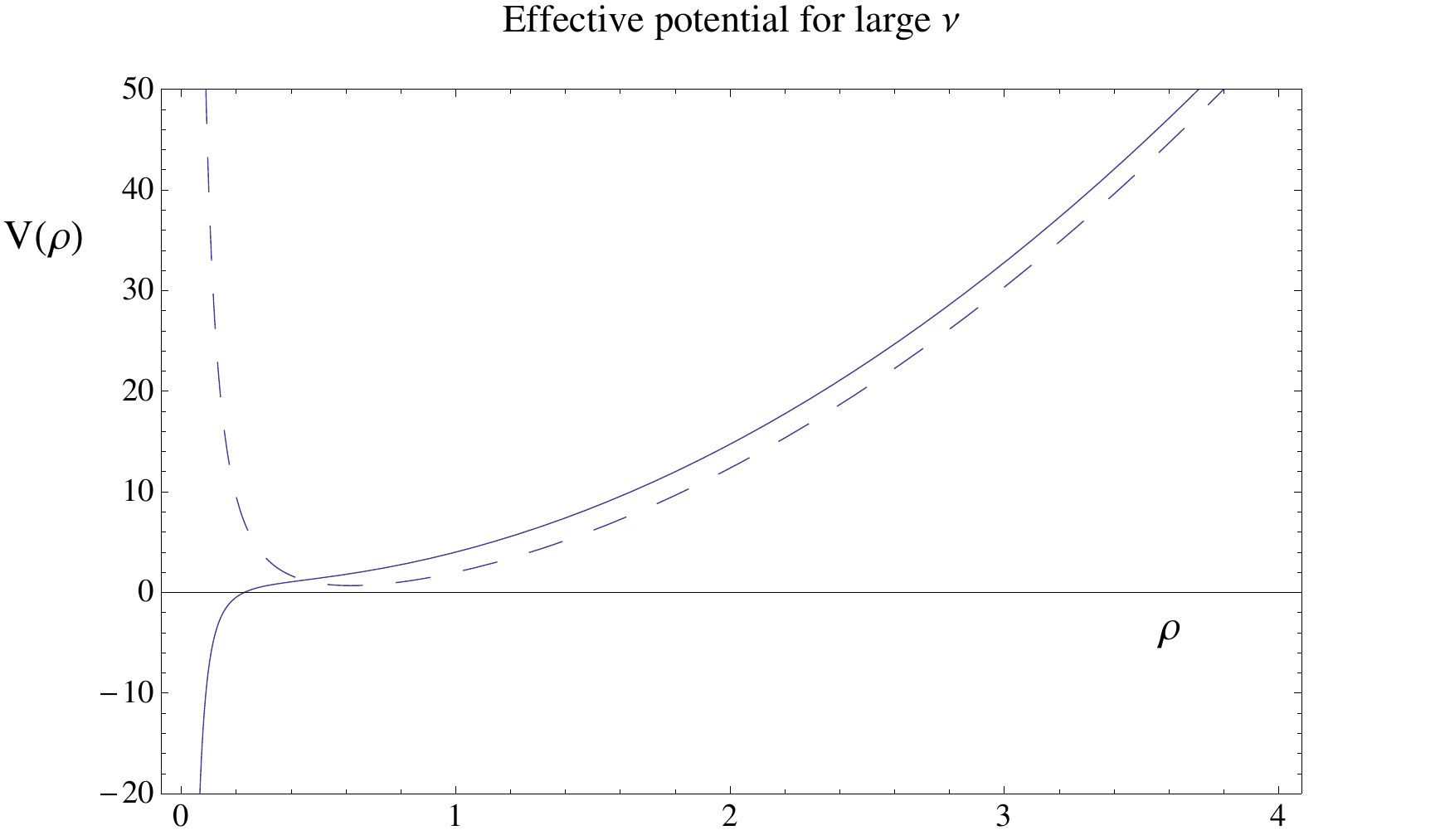}
\end{center}
\caption{Shape of the dimensionless effective potential $V(\rho)$ as a function
of the radial coordinate $\rho$.
(a) $\nu = 0.5$, $b = 0.1$; (b) $\nu = 0.5$, $b = 10.0$; (c) $\nu = 5.0$, $b = 1.0$.
In all figures the solid line corresponds to $m = 0$, and the dashed line to $m = 1$.}
\end{figure}

\section{The ground-state energy and probability current}

In the presence of the Coulomb interactions the system is not exactly solvable
and the numerical calculations are necessary (although analytical approximations
like WKB or dominant balance are readily available). To solve Eq. (\ref{TIprime})
we have used the standard Rayleigh-Ritz variational method \cite{Messiah}
with radial Gaussian functions $\rho^{(1/2) + m + k} \exp(-(1/2) \rho^{2})$,
$k = 0, 1, 2,...$, as the basis set.

The spectrum of the system is determined by two parameters $\nu$, $b$ and labelled
by the angular momentum quantum number $m$. For sufficiently small magnetic field
(i.e. sufficiently small parameter $\nu$), the ground state of the system is contained
in the manifold of states labelled by $m = 0$. With growing $\nu$, however,
the ground state can be associated with larger $m$.

In Fig. 2 we have shown the dependence of the ground-state energy on the
dimensionless strength
of the magnetic field for two values of $b$ and several values of the quantum
number $m$. It is clear that there exist values of the magnetic induction
for which the ground states becomes degenerate. More importantly, if the magnetic
field is sufficiently strong, the ground state corresponds to the non-zero
value of $m$. Clearly, this is a result of the interplay between the terms
in the Hamiltonian which are linear and quadratic in the magnetic potential.
As the latter approaches zero, the energy spectra for positive and negative $m$
cease to differ, of course.

\begin{figure}
\begin{center}
\includegraphics[width=7cm, height=6cm]{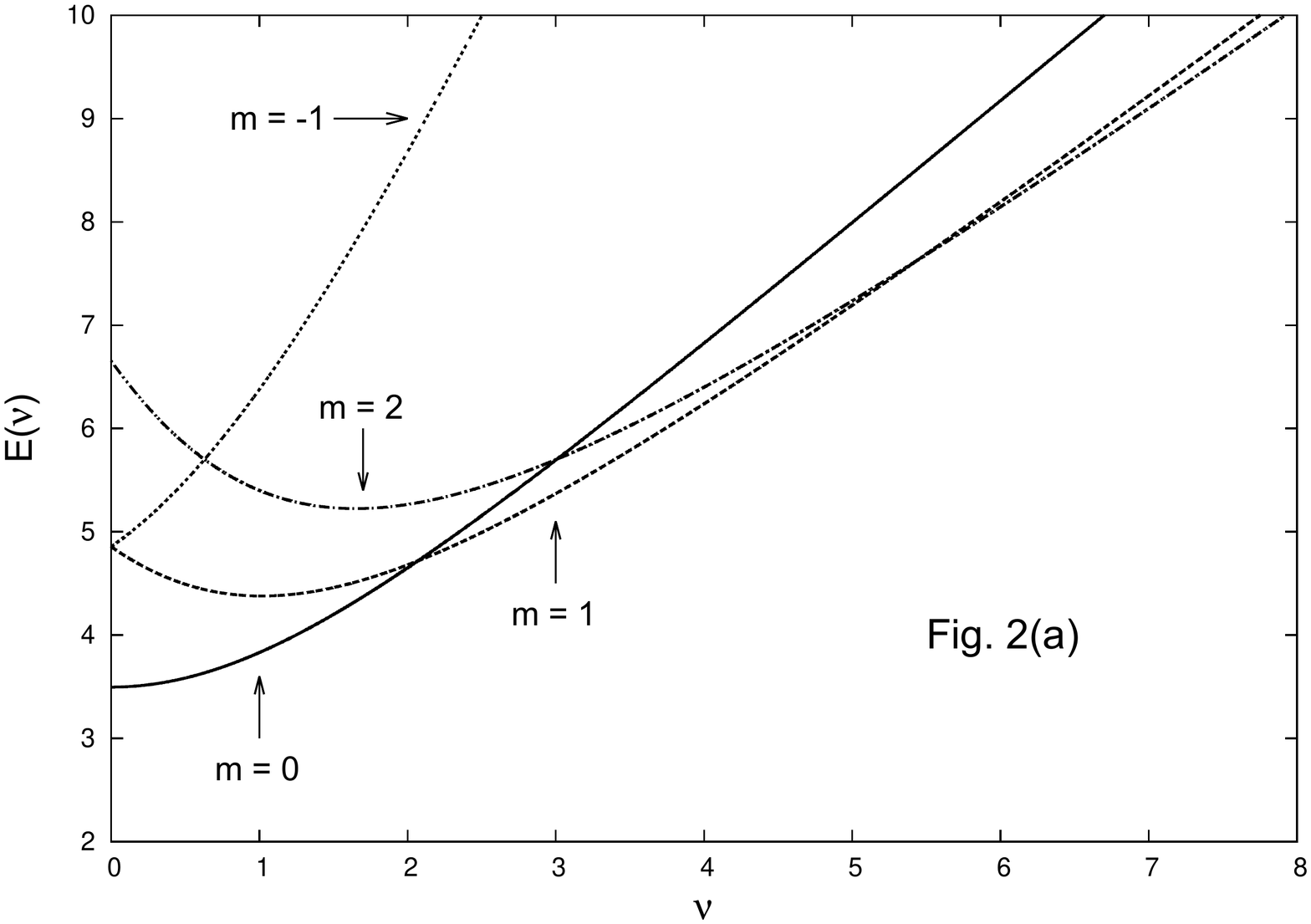}
\includegraphics[width=7cm, height=6cm]{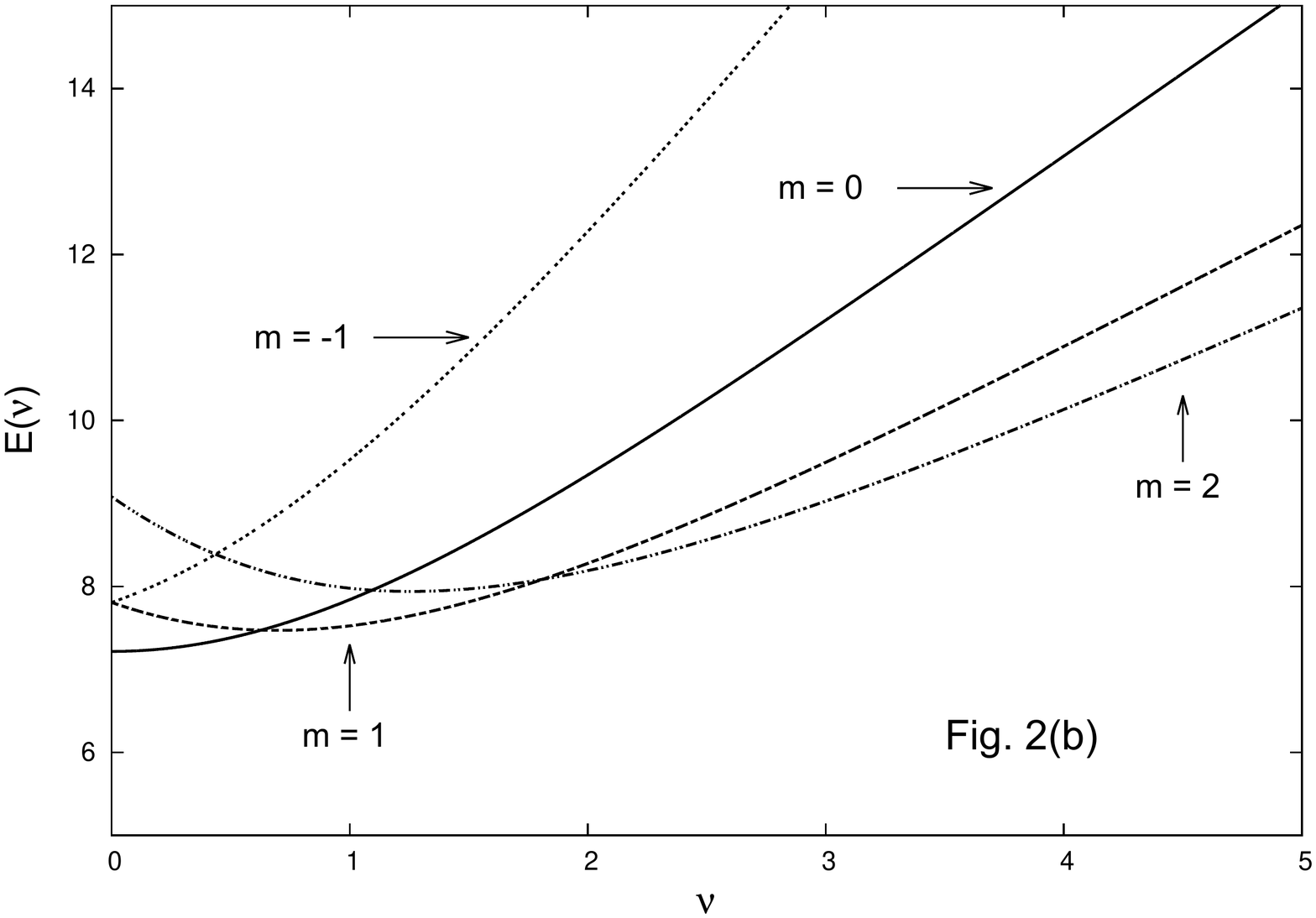}
\end{center}
\caption{Dependence of the ground-state energy on the scaled magnetic induction
(a) $b = 1.0$; (b) $b = 5.0$.
In all figures the solid line corresponds to $m = 0$, the dotted line to $m = 1$.
the short-dashed line to $m = 2$, and the long-dashed line - to $m = -1$.}
\end{figure}

Let us notice that no anticrossing appears at the points where various branches
of the spectrum are about to meet. The energy of the ground state is a non-differentiable
function of the magnetic induction.

As one might expect, a good approximation of the spectrum can be obtained
just by taking the expectation values of the Hamiltonian with the crude
apprxoximate wave functions with correct behavior at zero and infinity, i.e.

$$
\chi_{m} = \rho^{m+1/2} \exp(-(1/2) a \rho^{2}),
$$

where $a = \sqrt{1 + \nu^{2}/4}$.

In the context of the present paper, the probability current density in the stationary
states is of considerable interest. It is the current that gives adequate meaning to possible 
motion of the system, in the ground state in particular.
In radial coordinates, the only nonvanishing
component of the probability current density is the angular component,
${\bf j}(\rho, \phi) = (0, j(\rho))$ which does not depend on $\phi$. The natural
unit of the probability current is obtained from the natural unit of length
$\sqrt{\mu \omega_{t}/{\hbar}}$ and frequency $\omega_{t}$. Hence we define
the dimensionless current density ${\bf J}(\rho) = (0, J(\rho))$ by
${\bf j}(\rho) = \sqrt{\mu \omega_{t}/{\hbar}} \omega_{t} {\bf J}(\rho)$.
In Fig. 3 we have shown the behavior of $J$ as a function of its argument $\rho$ for
$b = 5.0$, $\nu = 1$, $m = 1$. For those values of $b$ and $\nu$
it is $m=1$ for which the ground state is achieved. Fig. 3(a) displays
the dependence of $J$ on $\rho$ while Fig. 3(b) shows the vector plot of
of ${\bf J}(\rho)$ in the $xy$-plane as computed in the ground state.

\begin{figure}
\begin{center}
\includegraphics[width=7cm, height=6cm]{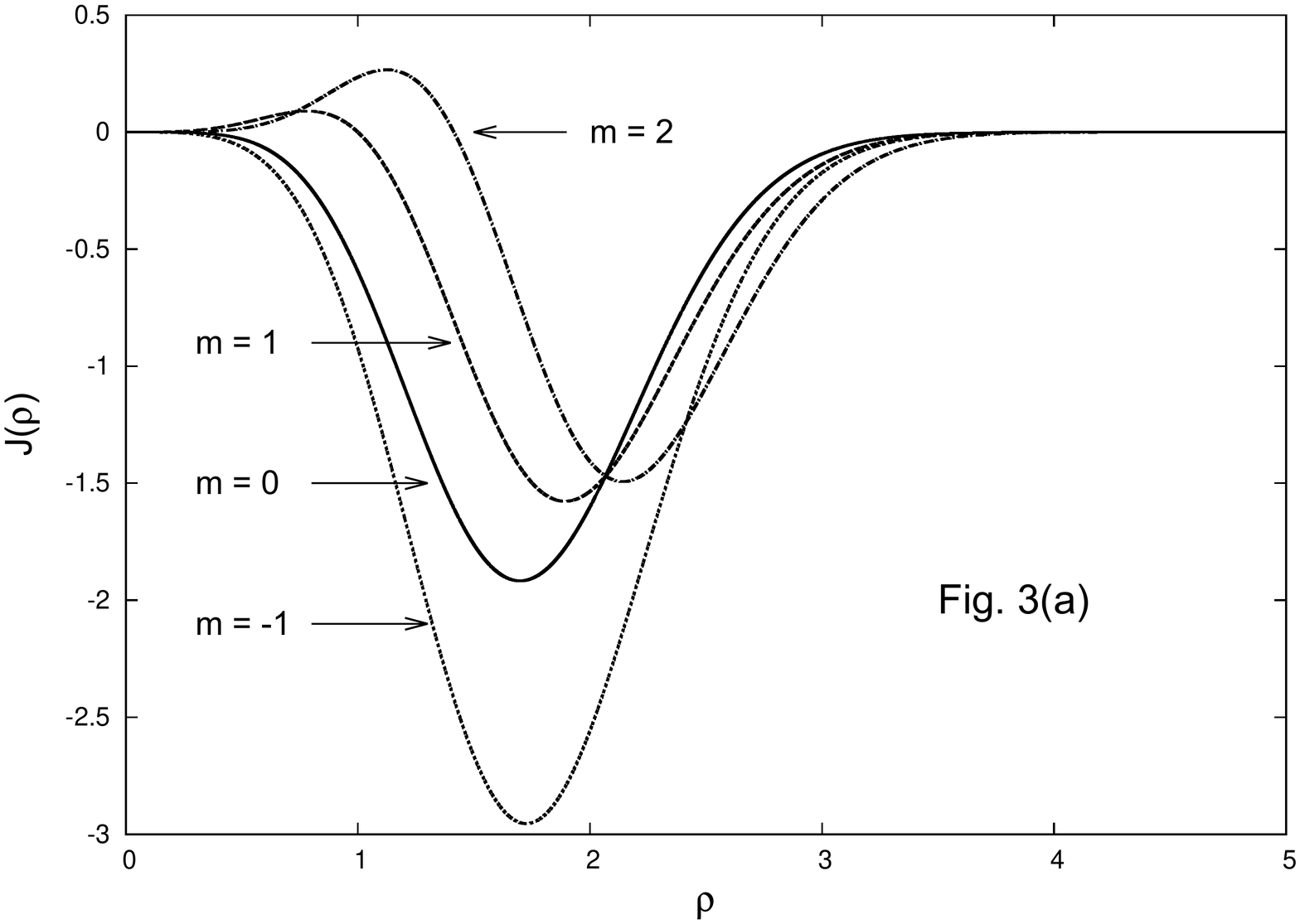}
\includegraphics[width=7cm, height=6cm]{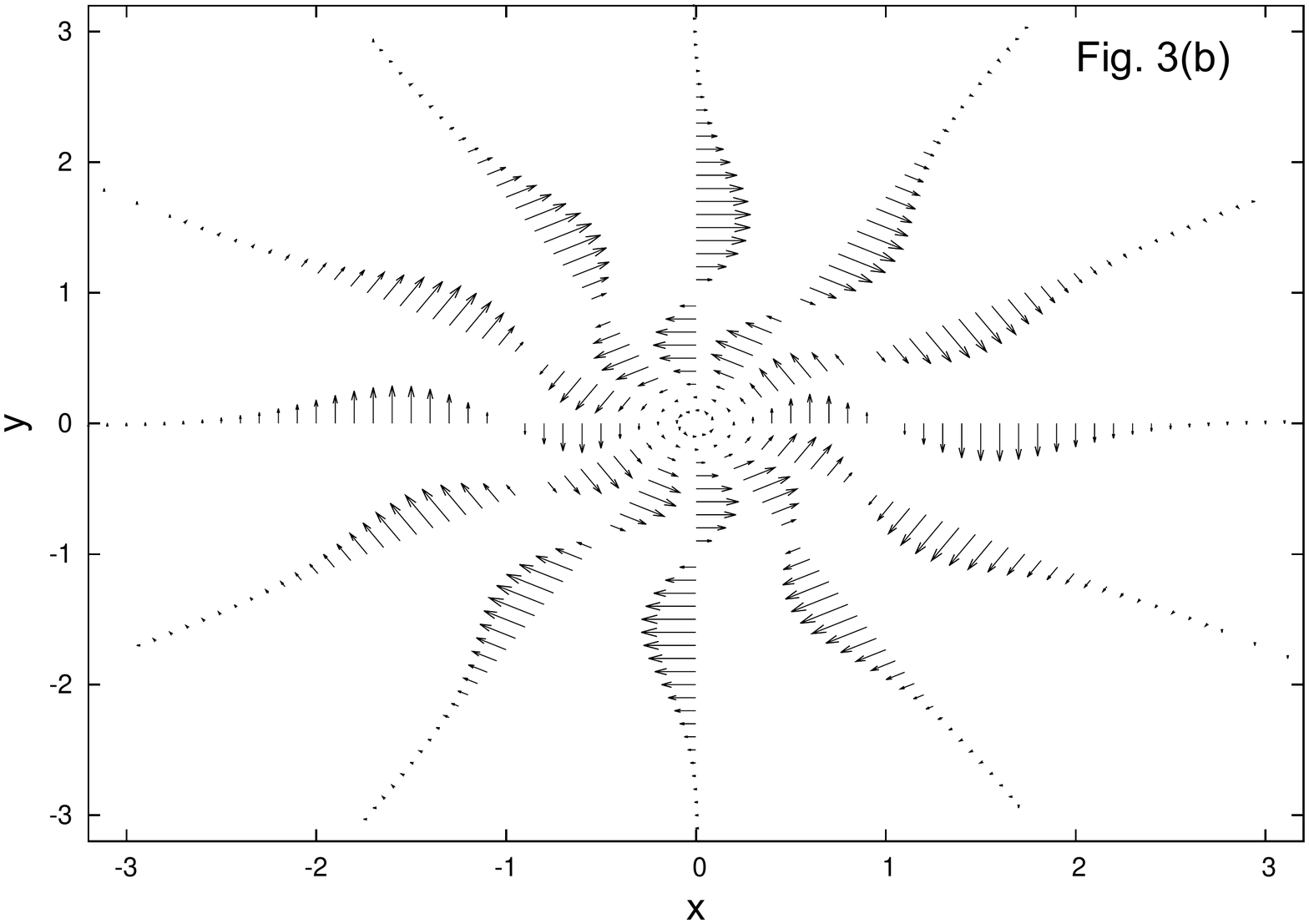}
\end{center}
\caption{Dependence of the probability current density on the position
in the $xy$ plane.
(a) dependence of the probability current density on the radial coordinate;
(b) vector plot of the current density as a functions of the position
in the plane.}
\end{figure}

Interestingly, the probability current density changes its sign with growing
$\rho$ for positive angular momentum.
The probability current itself (the density integrated over the whole plane)
does not vanish, however, in spite of the sign changes. Thus, the expectation value
of velocity (kinetic momentum) is non-zero in the ground state of the system.
This must be the case since the expectation
value of the radial coordinate does not vanish, hence the expectation value
of the vector potential is also non-zero. The fact that the expectation
value of velocity cannot vanish is especially obvious for the zero angular
momentum $m$: indeed, the azimuthal coordinate of the gradient of the wave
function is equal to zero, but the expectation value of the vector potential is not.

\begin{figure}
\begin{center}
\includegraphics[width=7cm, height=6cm]{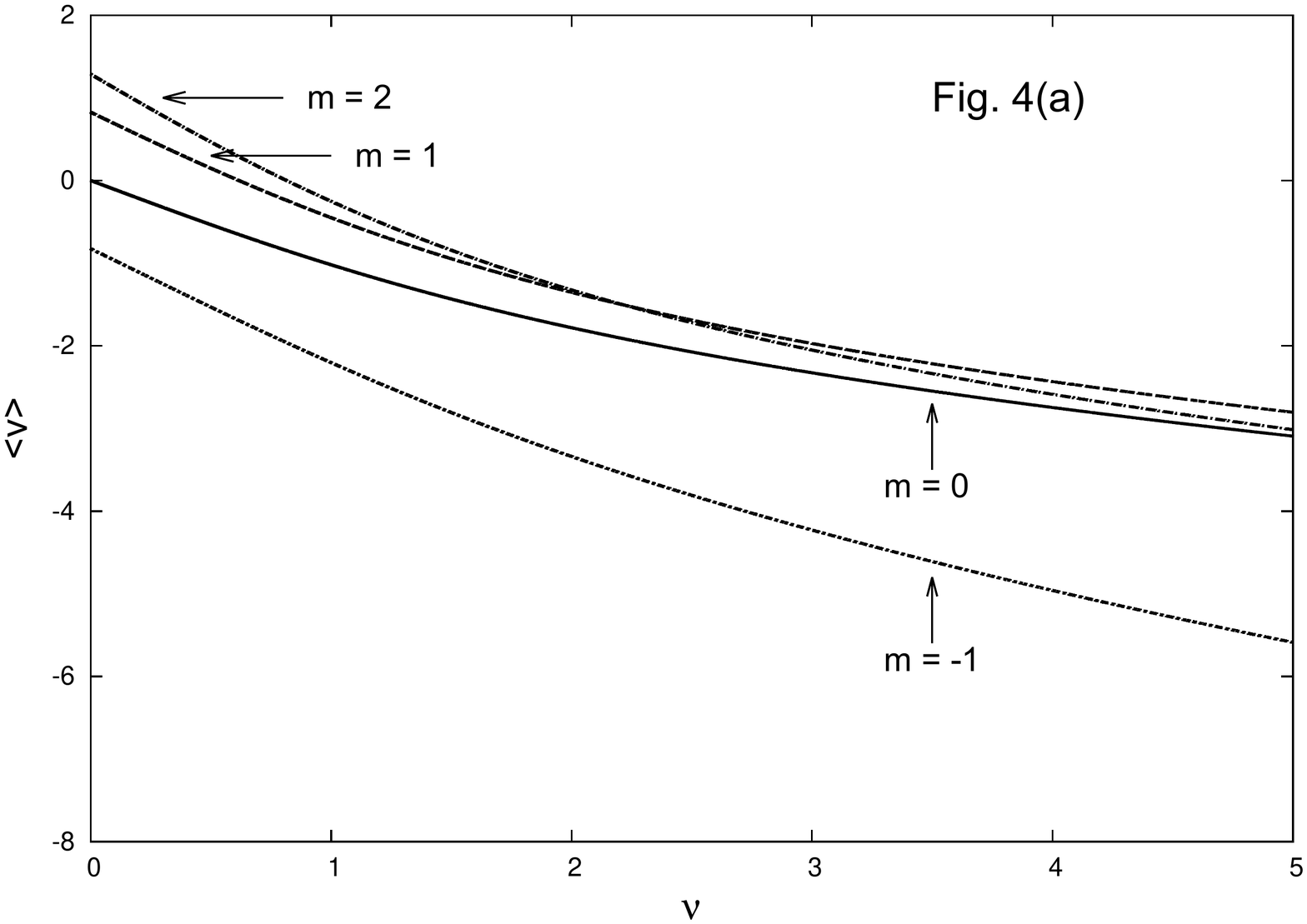}
\includegraphics[width=7cm, height=6cm]{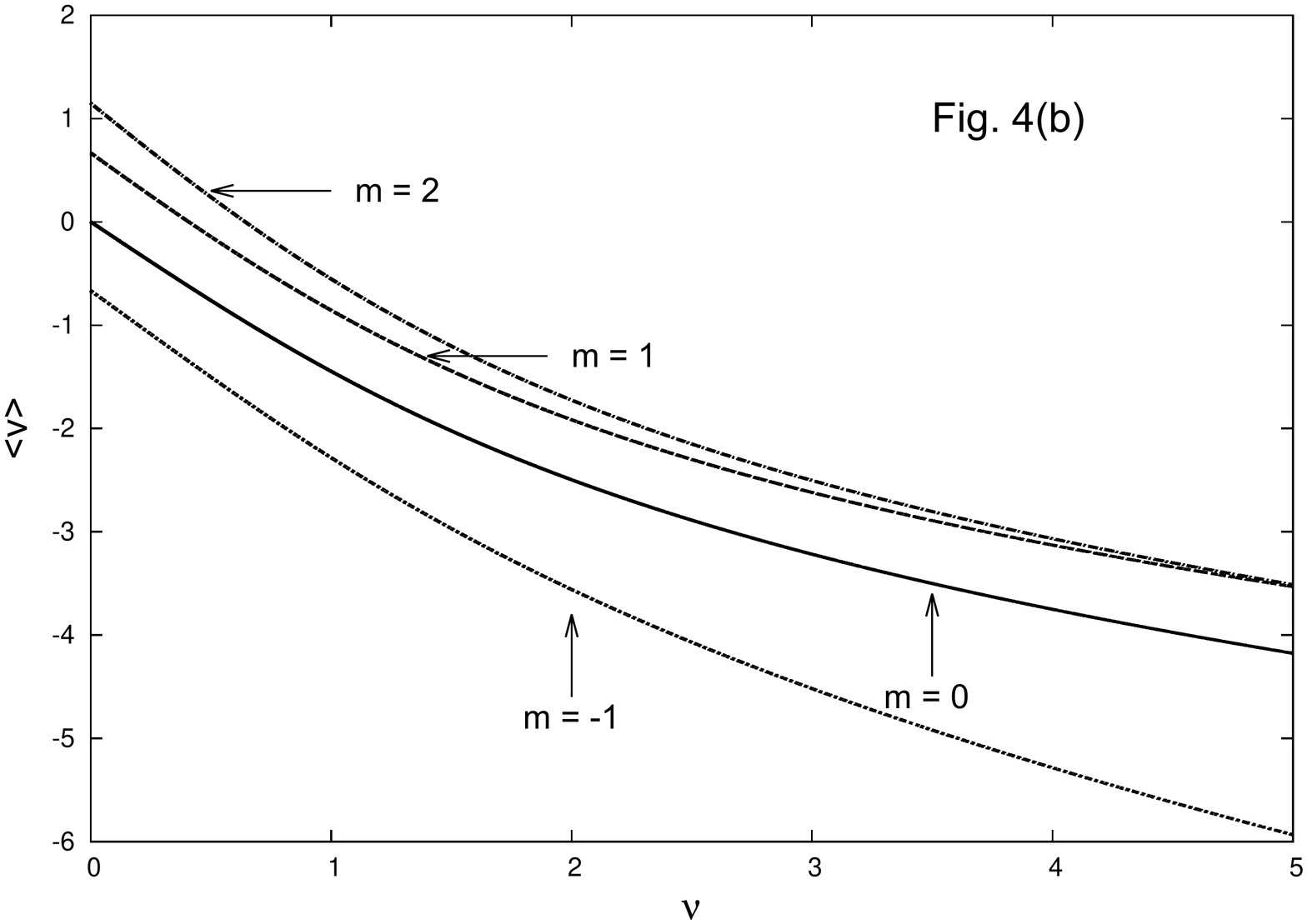}
\includegraphics[width=7cm, height=6cm]{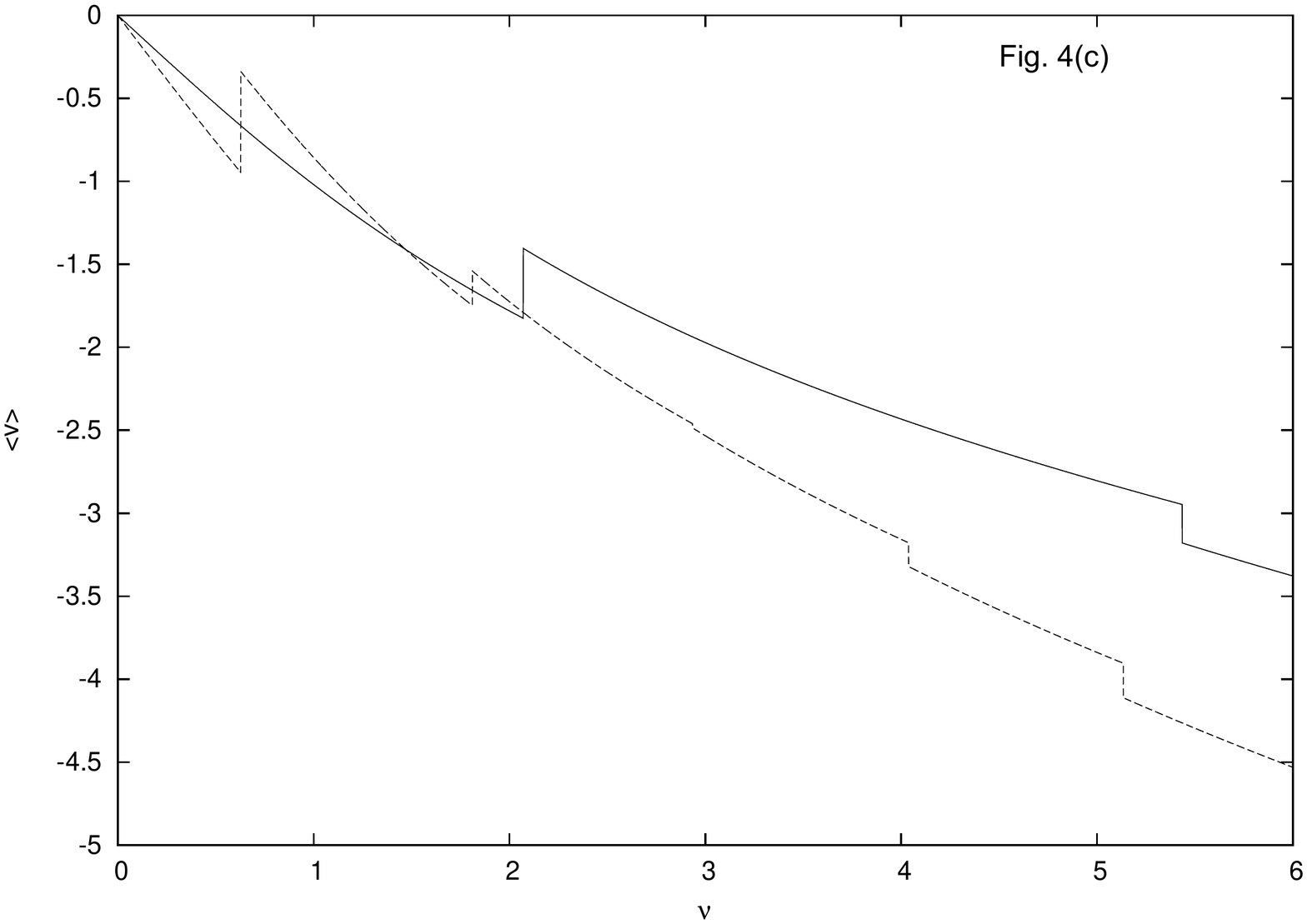}
\end{center}
\caption{Dependence of the expectation value of the (dimensionless) velocity
on the dimensionless magnetic field $\nu$.
(a) $b=1$; the solid line: $m = 0$, long-dashed line: $m = 1$, short-dashed line: $m = 2$,
dotted line: $m = -1$.
(b) the same as in (a) but for $b = 5$;
(c) dependence of the ground-state expectation value of velocity as a function
of the magnetic field $\nu$; change of the angular momentum corresponding to
the ground state as the magnetic field changes has been taken into account;
solid line: $b = 1$, dashed line: $b = 5$.}
\end{figure}

What we found very interesting is the behavior of the expectation value
of the velocity as dependent on the magnetic field. If Fig. 4(a) and 4(b) we have displayed
that dependence for several values of the angular momentum. Again, for positive values
of the angular momentum the expectation value of velocity changes its sign.
The magnetic field attempts to move the particle
in the positive direction of $\phi$. Thus, in order to "work out" the positive
angular momentum the particle must acquire negative (azimutal) velocity.
Fig. 4(c) illustrates the dependence of the ground-state expectation value of the velocity
as a function of the magnetic field $\nu$. It contains discontinuities because
different value of the angular momentum $m$ correspond to the ground-state energy
as the magnetic field grows as shown in Fig. 2.

\section{Dynamics of the wave packet}

In the next step in our considerations we have studied the dynamics of the wave function
which has initially been prepared as a Gaussian wave packet of the form:

\begin{equation}
\psi(\xi, \eta; 0) = N \exp(-a ((\xi - \xi_{0})^{2} + \eta^{2})).
\end{equation}

One can identify the following physical mechanisms which influence the dynamics
of the wave packet. The first is the harmonic force which, if present alone, would
simply yield the oscillations of the packet. The second one is the magnetic field
with twofold effect: the quadratic part introduces a new frequency to the system
which results in the ``breathing" of the wavepacket; the linear part generates
its overall rotation. Finally, there is the ``scattering" by the repulsive
Coulomb potential which leads to deformation of the packet as shown below.

The dynamics has been studied with the help of the standard split-operator
technique working in Cartesian coordinates. It has been convenient to work
in the dimensionless time $\tau = (1/2) \omega_{t} t$.
The time-evolution operator $U$ can be written in the form:

\begin{eqnarray}
U &=& \exp(-i h \tau) = \exp(-i h_{1} \tau) \exp(-i (h - h_{1}) \tau) = \nonumber \\
&=& \exp(-i h_{1} \tau) \exp(-i h_{2}) \tau) = U_{1} U_{2},
\end{eqnarray}

where $h = H/(\hbar \omega_{t})$, $h_{1} = i \nu (\xi \partial_{\eta} - \eta \partial_{\xi})$.
This results from the fact that the ${\bf A} \cdot {\bf p}$ term which generates
the rotations in the $x-y$ plane, commutes with the rest of the Hamiltonian.
Therefore, it has been convenient to work in the rotating frame $(\xi^{\prime}, \eta^{\prime})$
such that:

$$
\xi = \xi^{\prime} \cos(\tau) + \eta^{\prime} \sin(\tau)
$$

$$
\eta = -\xi^{\prime} \sin(\tau) + \eta^{\prime} \cos(\tau)
$$

The action of $U_{2}$ on $\psi$ has been computed according to:

\begin{eqnarray}
\psi(\xi, \eta, \tau + \Delta \tau) &\approx& \exp(-i (V_{2}/2) \delta \tau)
\cdot \nonumber \\
&\cdot& \exp((i/2)
(\partial_{\xi}^{2} + \partial_{\eta}^{2})\Delta \tau) \nonumber \\
&\cdot& \exp(-i (V_{2}/2) \Delta \tau)
\psi(\xi, \eta, \tau),
\end{eqnarray}

where $V_{2} = (1/2) (1 + 1/4 \nu^{2}) (\xi^{2} + \eta^{2}) + b/\rho$. The action
of the exponential of the kinetic term has been computed using two-dimensional fast Fourier
transformation and its inversion.

The snapshots of shape of the resulting wave functions have been displayed in Fig. 5.

\begin{figure}
\begin{center}
\includegraphics[width=3.1cm, height=3.1cm]{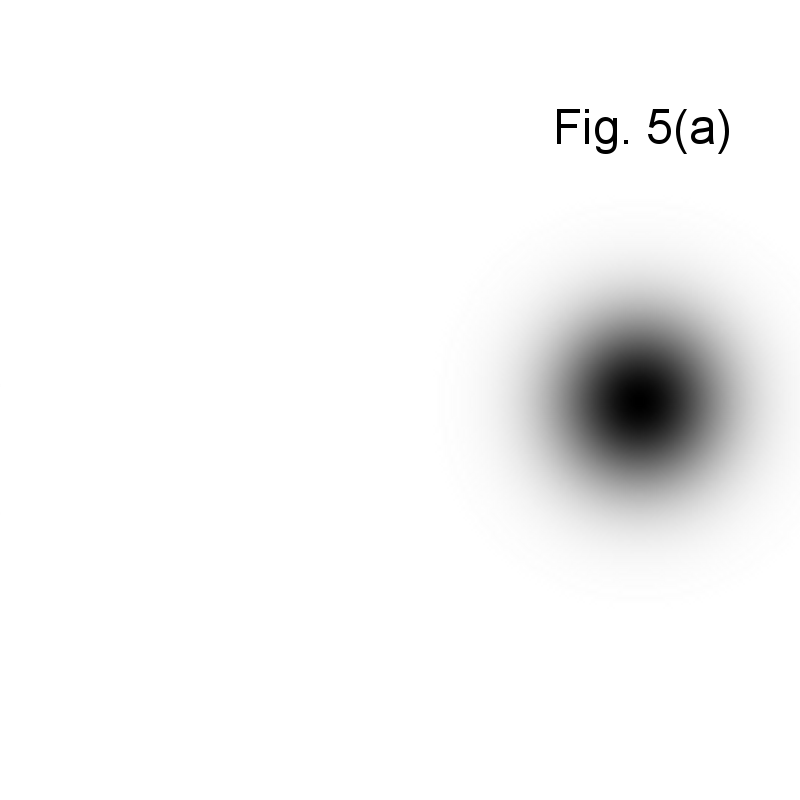}
\includegraphics[width=3.1cm, height=3.1cm]{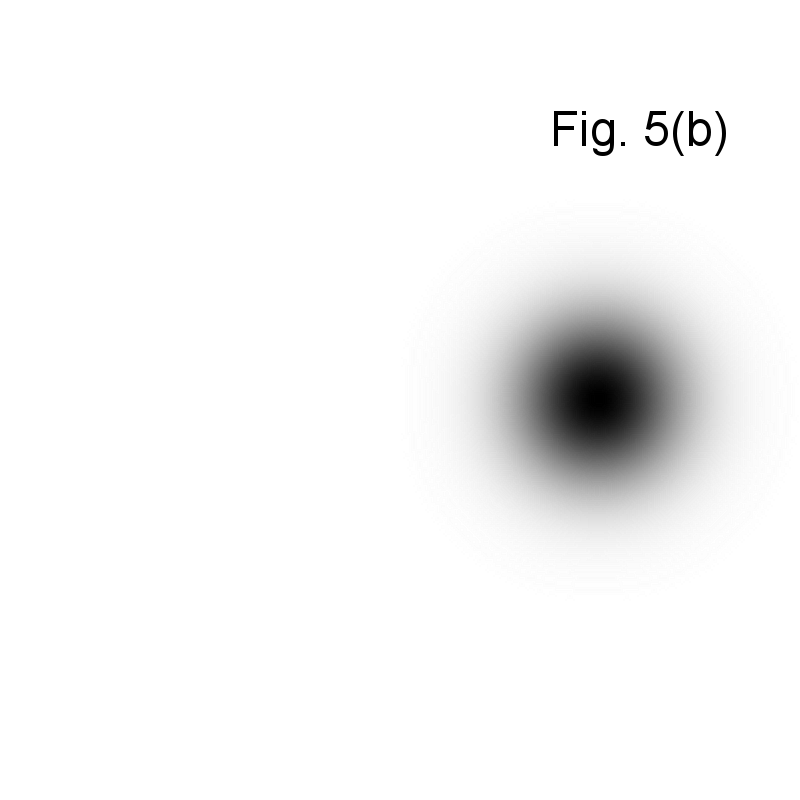}
\includegraphics[width=3.1cm, height=3.1cm]{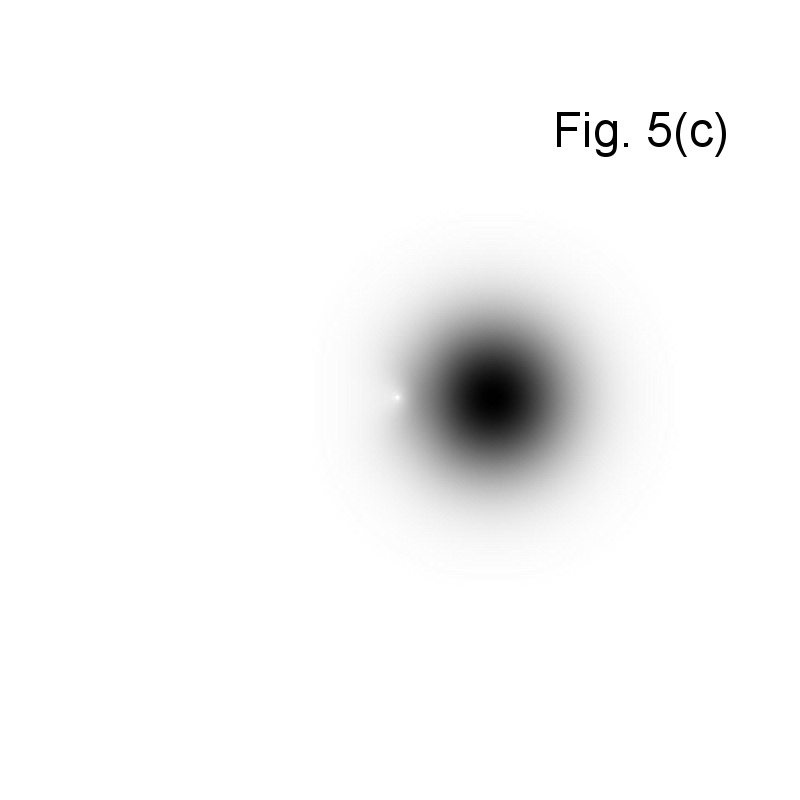}
\includegraphics[width=3.1cm, height=3.1cm]{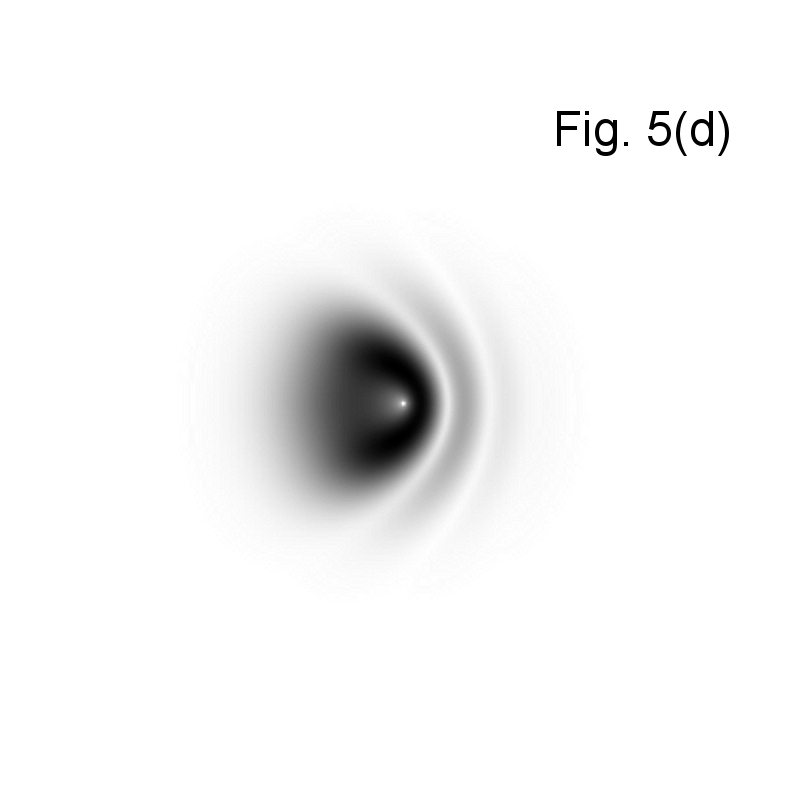}
\includegraphics[width=3.1cm, height=3.1cm]{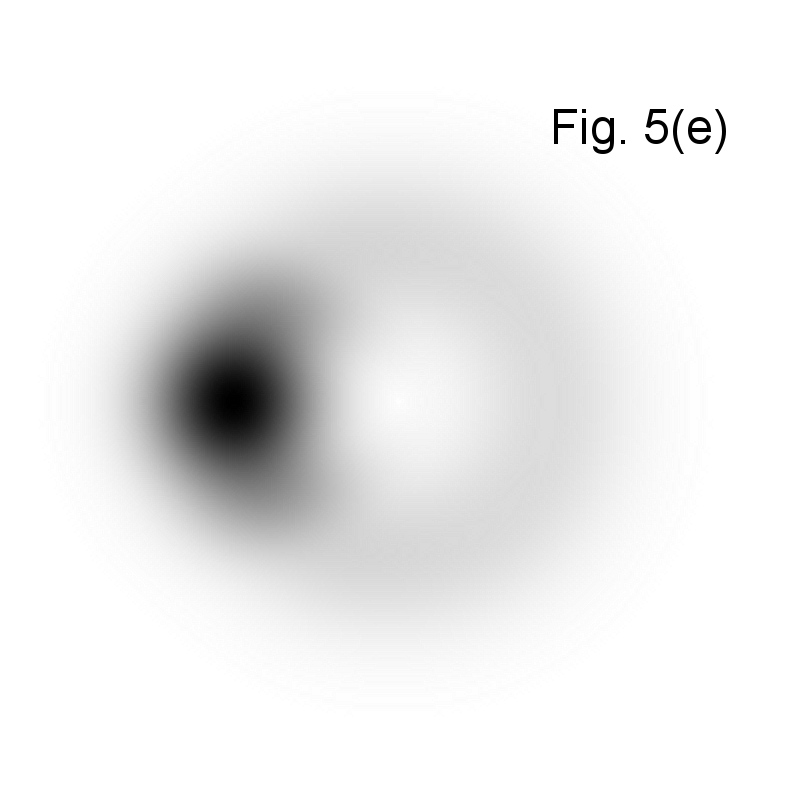}
\includegraphics[width=3.1cm, height=3.1cm]{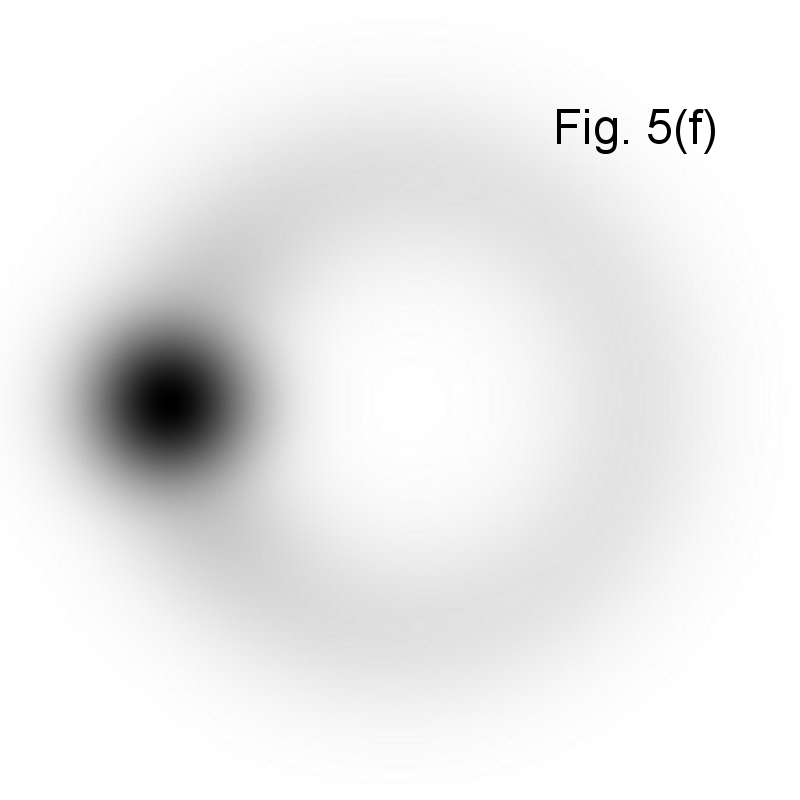}
\end{center}
\caption{Shaded contour plots of the time-dependent wave packets
in the rotating frame for $b = 1, \nu = 1$.
(a) Initial Gaussian wave packet localized at the point $\xi^{\prime} = 4,
\eta^{\prime} = 0$.
(b)-(e) Shapes of the wave packet at the time $\tau = j \pi/12$, $j=1,2,...,5$.
The darker regions corresponds to larger values of the modulus of wave function.}
\end{figure}

\setcounter{figure}{4}

\begin{figure}
\begin{center}
\includegraphics[width=3.1cm, height=3.1cm]{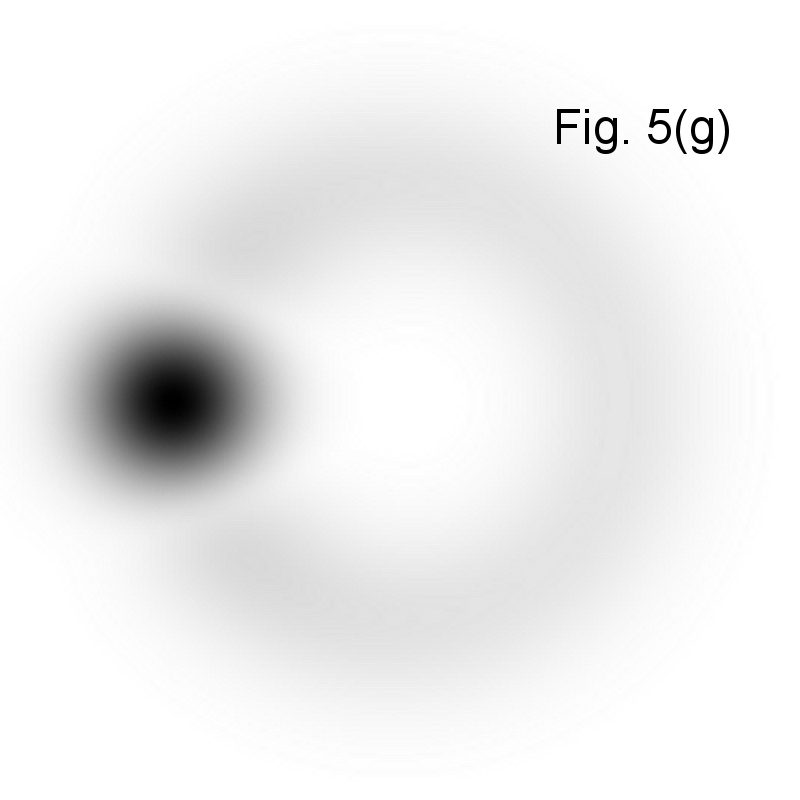}
\includegraphics[width=3.1cm, height=3.1cm]{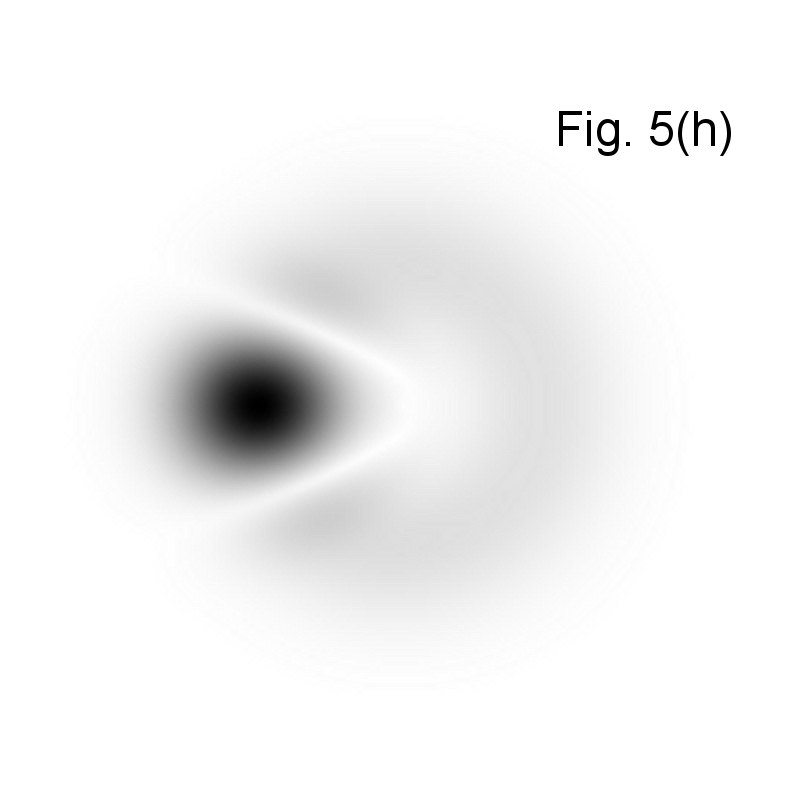}
\includegraphics[width=3.1cm, height=3.1cm]{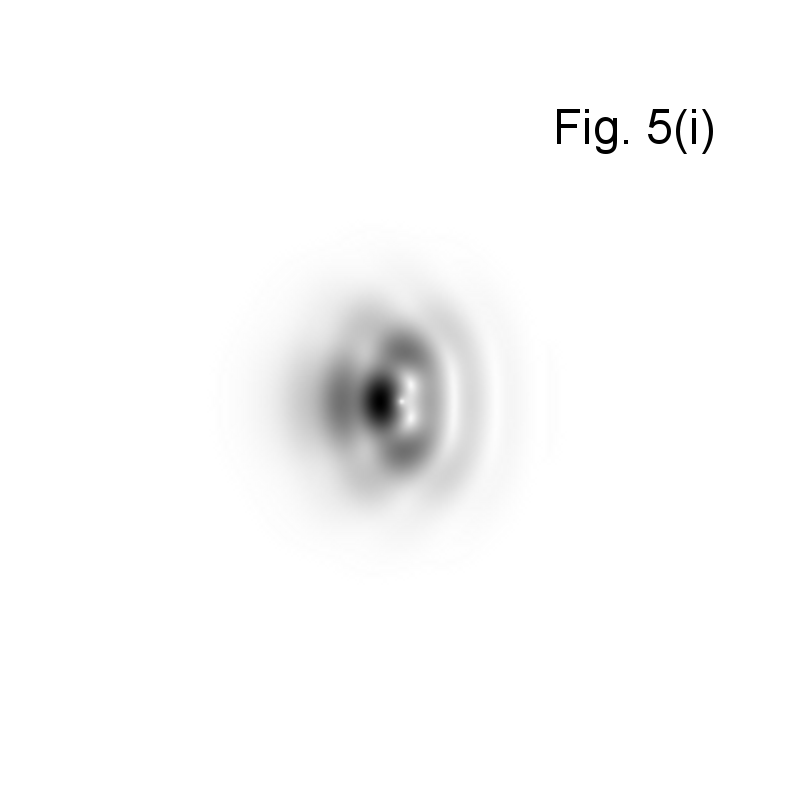}
\includegraphics[width=3.1cm, height=3.1cm]{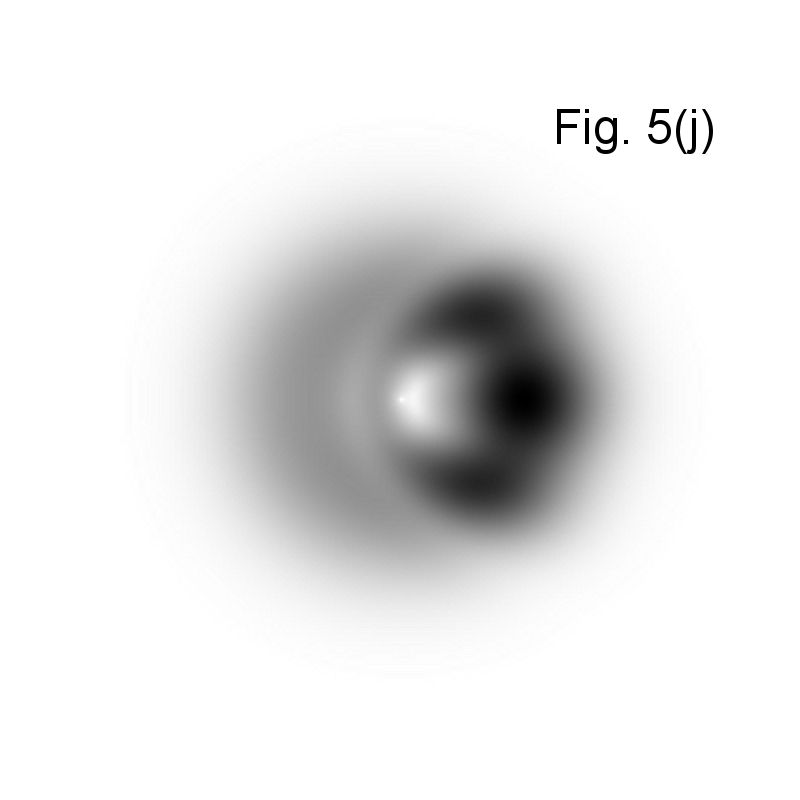}
\includegraphics[width=3.1cm, height=3.1cm]{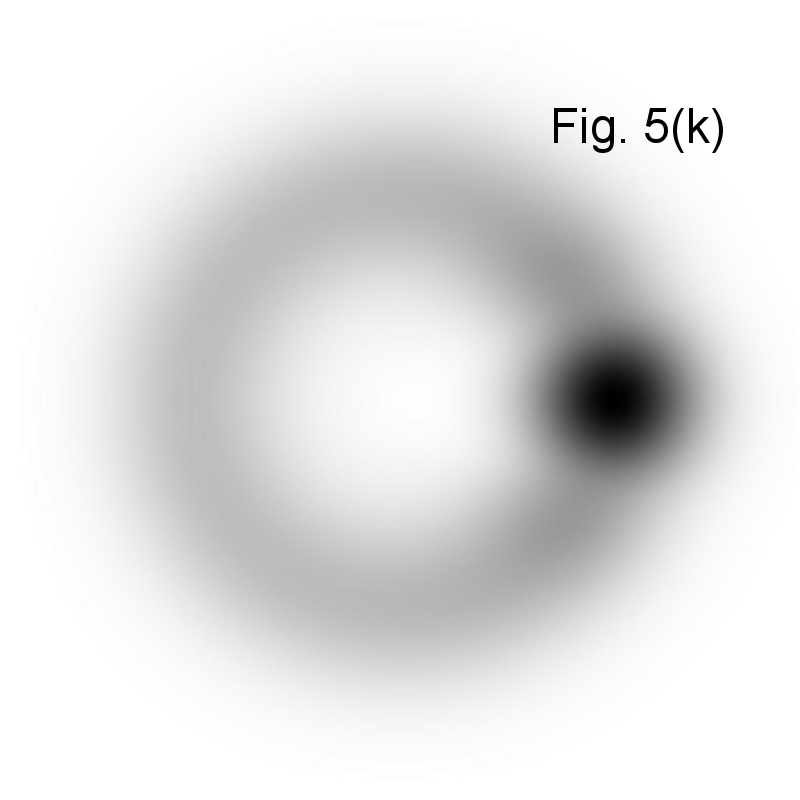}
\end{center}
\caption{(Continued) Shaded contour plots of the time-dependent wave packets
in the rotating frame for $b = 1$, $\nu = 1$.
(g)-(k) Shapes of the wave packet at the time $\tau = j \pi/12$, $j=6,7,...,10$.
The darker regions corresponds to larger values of the modulus of wave function.}
\end{figure}

As can be seen from the above figure, the (initially Gaussian) wave packet does
not return to its initial shape and becomes spread out in the azimutal coordinate 
due to the Coulomb ``scattering''.

\section{Concluding remarks}
We have also investigated whether the values of the magnetic induction leads to any
unusual characteristic in the ground state. Instead of the expansion in terms
of the basis sets, we have rather employed the dynamics in the imaginary time
as an alternative way to find the ground state.
However, we have not observed any spectacular changes in the imaginary time dynamics,
which might be associated with such a value of magnetic field which makes the ground
state degenerate.

Let us also notice that we have tested the system for adiabatic versus abrupt turning
on of the magnetic field, looking for any unusual behavior like phase
discontinuities. However, no such effects have been found.

To summarize, we have investigated a system of two identically charged particles
trapped in the harmonic potential with additional constant and homogeneous magnetic field.
The problem becomes effectively one-particle upon introduction of the relative coordinates.
The ground state of such a system can be degenerate for specific values of the magnetic
induction. Also, the ground state of the system can be that of non-vanishing eigenvalue
of the third component of the angular momentum. The ground-state expectation value
of velocity turns out to be non-zero. As a function of the magnetic field, it is discontinuous
at the points where the ground state becomes degenerate.
An interesting feature of the system is the sign change of the azimuthal component
of the probability current density in the ground state as a function of the radial coordinate.
We have also investigated the dynamics of the wavee packet. Due to the presence of the Coulomb
potential, the wavepacket partially loses its Gaussian shape and symmetry while retaining its coherence.
It becomes spread out in the ring trap.
No unusual behavior of the wavepacket have been observed while varying the magnetic field.

\end{document}